\documentclass[aps,prc,tightenlines,preprint,showpacs,groupedaddress]{revtex4}

\usepackage{graphicx}
\usepackage{units}
\usepackage{cancel}

\newcommand{\eq}[1]{Eq.~(\ref{#1})}
\newcommand{\be}{\begin{equation}}
\newcommand{\ee}{\end{equation}}
\newcommand{\bea}{\begin{eqnarray}}
\newcommand{\eea}{\end{eqnarray}}

\def\bfq{{\bf q}}

\begin{document}

\preprint{NT@UW-09-05}

\title{Connecting the Breit Frame to the Infinite Momentum Light Front
  Frame: How $G_E$ turns into $F_1$}

\author{Jared A. Rinehimer}
\author{Gerald A. Miller}
\affiliation{Department of Physics, University of Washington, Seattle, Washington 98195-1560, USA}

\date{\today}

\begin{abstract}
We investigate the connection between the Breit and infinite momentum frames and show 
that when the nucleon matrix element of the time component of the
electromagnetic current, which yields $G_E$ in the
Breit frame, is boosted to the infinite momentum (or light-front) frame, the quantity $F_1$ is
obtained.
\end{abstract}

\pacs{14.20.Dh, 13.40.Gp}

\maketitle


\section{Introduction\label{Intro}}
A tremendous amount of successful experimental effort has been
devoted to measuring the electromagnetic form factors of the nucleons (see the reviews 
\cite{Gao:2003ag,HydeWright:2004gh,Perdrisat:2006hj,Arrington:2006zm}).
The experiments have achieved unprecedented accuracy, but the
interpretation of the form factors in terms of charge or magnetization
densities has been clouded by the need to understand the relativistic
motion of the target as a whole and of the ultrarelativistic motion of
the light $u$ and $d$ quarks moving within.

The standard interpretation follows from the fact 
 that the nucleon helicity-flip matrix element
of the time component of the electromagnetic current density, when
 evaluated in the  Breit frame [in which initial momentum of the
 nucleon is antiparallel to that of the incident virtual photon ($\bfq$)],
 yields $G_E$. Thus $G_E$ is the matrix element of the charge density
 under the stated  conditions. This connection has been used to imply
 that the charge density is the three-dimensional Fourier transform of
 $G_E$. However, the initial and final nucleons have different momenta
 and therefore have different wave functions. The separation between
 relative and center of mass variables that occurs under
 nonrelativistic dynamics does not occur for rapidly moving
 constituents or targets. Thus the initial and final wave functions
 are related by a boost that generally depends on interactions.

This difficulty can be surmounted by using an infinite momentum frame
analysis, with the Drell-Yan condition that
$q^+=(q^0+q^3)=0$, where the infinite momentum component of the nucleon is in the  $z$
direction. In this case, one obtains a model-independent transverse
density that is the two-dimensional Fourier transform of the $F_1$
form factor \cite{soper1,mbimpact,diehl2, MillerNeutron, Carlson:2007xd}.

One consequence of using this model-independent formalism
is that the central value of the transverse charge density of
the neutron is negative \cite{MillerNeutron}, 
in seeming contradiction to the long held
notion that the center of the neutron is positively charged. This
contradiction may arise simply from working in the infinite momentum frame
to extract the transverse density.
 Our purpose here is to examine the connection between using the 
 Breit and infinite momentum frames to compute the matrix element of
 the charge density operator. 

We generalize the usual Breit frame formalism to include the use of
arbitrary spin directions for the initial and final states in
Sec.~II.  These states are related to those defined by the use of
light-cone spinors in Sec.~III, and the use of the infinite momentum
frame is discussed in Sec.~IV. The remaining section is used for a
brief summary and discussion.

\section{Generalized Breit Frame Formalism} 

The form factors are defined by the matrix element of the electromagnetic current operator $J^{\mu}(x)$ as
\begin{equation}\label{ff}{\langle}f|J^{\mu}(0)|i{\rangle}= \nonumber
\bar{u}_f(p')[{\gamma}^{\mu}F_1(Q^2)+i{\frac{{\sigma}^{{\mu}{\nu}}q_{\nu}}{2M}}F_2(Q^2)]u_i(p)
, \end{equation} 
where $q={p'}-p$ is the momentum transfer, which is space-like such
that $-q^2 \ {\equiv} \ Q^2 >0 $, $f$ 
indicates final state, and $i$ indicates initial state. 
The Sachs form factors \cite{MillerNeutron}, defined as
\begin{eqnarray}
G_E(Q^2)&=&F_1(Q^2)-\frac{Q^2}{4M^2}F_2(Q^2)\nonumber\\
G_M(Q^2)&=&F_1(Q^2)+F_2(Q^2)
\end{eqnarray} have the common interpretation of being related to
the charge 
and magnetization density of the nucleon, respectively. 

Consider the Breit frame (Frame A), with a nucleon with an incoming
four-momentum of $p_A^{\mu}=(E_0,p_0,0,0)$ 
and an outgoing four-momentum of ${p'}_A^{\mu}=(E_0,-p_0,0,0)$, leading  to
$q^\mu={p'}_A^\mu-p_A^\mu=(0,-2p_0,0,0)$ and $-q^2=Q^2=4p_0^2$. The
relevant Dirac spinors for this Breit frame analysis can be written as
\begin{equation}  u_{{\pm}x}(p_A)=\frac{1}{\sqrt{E_0+M}}(\cancel{p}_A+M)\left( 
\begin{array}{l} 
\chi_{{\pm}x} \\ 
0
\end{array}\right)
\end{equation}
and
\begin{equation}  \bar{u}_{{\pm}x}(p_A)=\frac{1}{\sqrt{E_0+M}}(\chi_{{\pm}x}^\dagger \ 0)(\cancel{p}_A+M), 
\end{equation}
where $\chi_{\pm{x}}=\nicefrac{1}{\sqrt{2}}( 1, \pm 1) $ is the
two-component spinor in the ${\pm}x$ direction, and  the normalization is
$\bar{u}(p)u(p)=2M$. 

We  compose the incoming state as a yet-to-be-determined state 
\begin{equation}
u_{i_A}(p_A)=a\ u_{+x}(p_A)+b\ u_{-x}(p_A), 
\end{equation}
and likewise  the final state is
\begin{equation}
{\bar{u}}_{f_A}(p'_A)=c\ \bar{u}_{+x}(p'_A)+d\ \bar{u}_{-x}(p'_A), 
\end{equation} 
with the constraints
\begin{eqnarray}
{|a|}^2+{|b|}^2&=&1 \nonumber \\
{|c|}^2+{|d|}^2&=&1, \label{norm}
\end{eqnarray}
maintaining the normalization condition.

When the current of Eq.~(\ref{ff}) is evaluated between these initial
and final states, one finds
\begin{eqnarray}
{\langle}f_A|J^{0}(0)|i_A{\rangle}&=&2M(ac+bd)\ G_E(Q^2)\label{first} \\
{\langle}f_A|J^{1}(0)|i_A{\rangle}&=&0 \\
{\langle}f_A|J^{2}(0)|i_A{\rangle}&=&-2i(ad+bc)p\ G_M(Q^2) \\
{\langle}f_A|J^{3}(0)|i_A{\rangle}&=&2(ad-bc)p\ G_M(Q^2). 
\end{eqnarray}

We see that it is possible to achieve a matrix element of the $J^0$
component that is proportional to $G_E(Q^2)$ by using a multitude of sets of
initial and final states in the Breit frame, not simply the typical
helicity-flip elements. This is why we use the term ``generalized"
Breit frame formalism.

We shall see that 
a convenient choice of the coefficients of the vectors is to use
\begin{eqnarray} 
a=\frac{\sqrt{E_0+M}}{\sqrt{2E_0}},\; 
b=\frac{-\sqrt{E_0-M}}{\sqrt{2E_0}},\; 
c=\frac{\sqrt{E_0+M}}{\sqrt{2E_0}},\;
d=\frac{\sqrt{E_0-M}}{\sqrt{2E_0}}.\label{choice}
\end{eqnarray} 
This choice has the property that, for large values of $Q^2$ (or
$E_0$),the initial and final states correspond to spins in the
$-z$ and  $+z$ directions. 
Moreover, \eq{first} becomes
\bea
{\langle}f_A|J^{0}(0)|i_A{\rangle}&=&{2M^2\over E_0}\
G_E(Q^2).\label{second}
\eea
The use of \eq{choice}
yields  the spinors
\begin{eqnarray}
&u_{i_A}(p_A)=\frac{1}{2\sqrt{E_0}}\left(
\begin{array}{c}
E_0+M-p_0 \\
E_0+M+p_0 \\
E_0-M+p_0 \\
-E_0+M+p_0
\end{array}\right)& \nonumber  \\ 
&\bar{u}_{f_A}(p'_A) = \frac{1}{2\sqrt{E_0}} \left( \begin{array}{c c c c} E_0+M+p_0, & E_0+M-p_0, & -E_0+M+p_0, & E_0-M+p_0 \end{array} \right),& \label{breit1}
\end{eqnarray}
Next we boost these in the $z$ direction to a new frame, so the incoming and outgoing states both have some momentum $p_z$ in the $z$ direction. This is achieved by boosting the spinors with the appropriate boost matrix, given by the formula \cite{Bjorken}
\begin{equation} \label{boost}
S=\textrm{exp}(-\frac{i\omega_i}{2}\sigma_{0i})=\frac{1}{\sqrt{2E_0}}\left(
\begin{array}{c c c c}
\sqrt{E+E_0} & 0 & \sqrt{E-E_0} & 0 \\
0 & \sqrt{E+E_0} & 0 & -\sqrt{E-E_0} \\
\sqrt{E-E_0} & 0 & \sqrt{E+E_0} & 0 \\
0 & -\sqrt{E-E_0} & 0 & \sqrt{E+E_0}
\end{array}\right),
\end{equation}
where {\bf v} is the boost velocity
$\mathbf{v}=\frac{-p_z}{E}\hat{z}$, $E=\sqrt{{p_z}^2+{p_0}^2+M^2}$ is
the energy of the initial and final states after boost, and the
rapidity
$\omega_i=\mathrm{tanh}^{-1}(|\mathbf{v}|)\frac{v_i}{|\mathbf{v}|}$.
After this boost,  $q^\mu$ remains unchanged while $p_A^\mu$ of the
incoming state is boosted to $p_{B}^\mu=(E,p_0,0,p_z)$. Likewise,
${p'}_{B}^{\mu}=(E,-p_0,0,p_z)$.  We denote this frame as Frame B.
The incoming Breit frame spinor
Eq.~(\ref {breit1}), 
when boosted with the matrix $S$ (\ref{boost}), is
\begin{eqnarray} 
 S \, u_{i_A}(p_A)
&=&  
\frac{1}{2\sqrt{p^+_B}}
\left(\begin{array}{c}
p_{B}^++M-p_0 \\
p_{B}^++M+p_0 \\
p_{B}^+-M+p_0 \\
-p_{B}^++M+p_0
\end{array}\right)\label{finspin}
\end{eqnarray}
where $p^{\pm}=p^0\pm p^3$ for any four-momentum $p^\mu$, 
and we have used the identity
$p_{B}^+p_{B}^-=E^2-{p_z}^2={p_0}^2+M^2={E_0}^2$. Note also that
$q^+=q^0+q^3=0$ , a condition that can be  chosen for any space-like virtual momentum transfer.

\section{Light Cone Spinors}
We can construct the ``light-cone" spinors 
\cite{Kogut:1969xa,Lepage:1980fj,Frederico:1991vb}
using the formula 
\begin{equation}  \label{def}u_{{\uparrow}x}(p)=\frac{1}{\sqrt{2p^+}}(\cancel{p}+M)\gamma^+\left( 
\begin{array}{l} 
\chi_{+x} \\ 
0
\end{array}\right),
\end{equation}
with an analogous definition for $u_{{\downarrow}x}$, and
\begin{equation}  \bar{u}_{{\uparrow}x}(p')=\frac{1}{\sqrt{2p^+}}(\chi_{+x}^\dagger \ 0){\gamma}^+(\cancel{p'}+M),
\end{equation}
with $\gamma^+=\gamma^0+\gamma^3$.
Using these definitions, we obtain the explicit representations
\begin{equation}
u_{\uparrow x}(p_{B})=\frac{1}{2\sqrt{p_{B}^+}}\left(\begin{array}{c} p_{B}^+ + M - p_0 \\ p_{B}^+ + M + p_0 \\ p_{B}^+ - M + p_0 \\ -p_{B}^+ +M + p_0 \end{array} \right),\hspace{10 pt} u_{\downarrow x}(p_{B})=\frac{1}{2\sqrt{p_{B}^+}}\left(\begin{array}{c} p_{B}^+ + M + p_0 \\ -p_{B}^+ - M + p_0 \\ p_{B}^+ -M - p_0 \\ p_{B}^+ -M + p_0 \end{array} \right), 
\label{add}\end{equation}
and
\begin{eqnarray} \nonumber
\bar{u}_{\uparrow x}({p'}_{B})=\frac{1}{2\sqrt{p_{B}^+}}\left(\begin{array}{c c c c} p_{B}^+ + M + p_0, & p_{B}^+ + M - p_0, & -p_{B}^+ + M + p_0, & p_{B}^+ -M + p_0 \end{array} \right),\\ 
\bar{u}_{\downarrow x}({p'}_{B})=\frac{1}{2\sqrt{p_{B}^+}}\left(\begin{array}{c c c c} p_{B}^+ + M - p_0, & -p_{B}^+ - M - p_0, & -p_{B}^+ +M - p_0, & -p_{B}^+ +M + p_0 \end{array} \right).
\end{eqnarray}
The Breit frame spinors (\ref {breit1}), when boosted with the matrix
$S$ (\ref{boost}), can now be decomposed in this basis. It is clear from
\eq{finspin} and \eq{add} that 
\begin{equation}
S\,u_{i_A}(p_A)=u_{\uparrow x}(p_{B}).
\end{equation}
Boosting on $\bar{u}_{f_A}(p'_A)$ from Eq. (\ref{breit1}) gives
\begin{equation}
\bar{u}_{f_A}(p'_A)\rightarrow \bar{u}_{f_A}(p'_A) \gamma_0 S^\dagger \gamma_0=\bar{u}_{\uparrow x}(p'_{B}).
\end{equation}
Additionally, examining Eq. (\ref{breit1}) further, we see that in the Breit frame $p_A^+=p_A^0+p_A^3=E_0$  and that we can write the spinors in Eq. (\ref{breit1}) as 
\begin{eqnarray}
u_{i_A}(p_A)=\frac{1}{2\sqrt{p_A^+}}\left(\begin{array}{c} p_A^+ + M - p_0 \\ p_A^+ + M + p_0 \\ p_A^+ - M + p_0 \\ -p_A^+ +M + p_0 \end{array}\right) = u_{\uparrow x}(p_A) \nonumber \\
\bar{u}_{f_A}(p'_A)=\bar{u}_{\uparrow x}(p'_A).
\end{eqnarray}
Thus the 
choice of $a,b,c,d$ of \eq{choice} corresponds to using
``light-cone" spinors in the Breit frame (Frame A) that are  boosted
 to the ``light-cone" spinors in the boosted frame (Frame B). 

\section{Evaluation in the infinite momentum frame (IMF)} 
In the infinite momentum frame the charge density operator $J^0$
becomes $J^+=J^0+J^3$. This is obtained using a Lorentz transformation
with the velocity taken to be infinitesimally close to unity and
noting that the ``$\gamma$'' factor is absorbed into the change in the
$z$ coordinate \cite{ls}.
 
We  evaluate the  matrix element of $J^+$ in Frame B to find 
\begin{equation}
{\langle}f_{B}|J^{+}(0)|i_{B}{\rangle}=2 p_{B}^+ F_1(Q^2)  \label{jplus}.
\end{equation}
Note that this relationship is independent of the boost parameter
$p_z$ and remains the same when we take the IMF limit. However, it is
only in the IMF that the operator $J^0$ becomes $J^+$.
If we boost to the IMF with infinite momentum $p_z$, helicity spinors are now defined with the spin aligned along the $z$ direction, because the momentum of the spinor in the $x$ direction is negligible compared to the large $z$ momentum. Because $u_{\uparrow x}(p_{B})=\frac{1}{\sqrt{2}}\left(u_{\uparrow z}(p_{B})+u_{\downarrow z}(p_{B})\right)$ as seen from the definition in Eq. (\ref{def}), we see that a matrix element in the IMF frame corresponding to a two-dimensional charge density can be formed with a linear combination of IMF helicity spinors, instead of only the expected helicity-non-flip matrix elements.

The results \eq{second} and \eq{jplus} are our central results. They
show that a Breit frame matrix element of $J^0$ yielding $G_E$ is converted to
a matrix element of $J^+$ that yields $F_1$ in the IMF.

Furthermore, we can attempt to begin with solely helicity spinors in
the IMF 
and determine what matrix elements 
they boost back to in the Breit frame.  It is also true that
\begin{eqnarray}
u_{\updownarrow z}(p_A) \rightarrow S\,u_{\updownarrow z}(p_A)=u_{\updownarrow z}(p_{B}) \nonumber \\
\bar{u}_{\updownarrow z}(p'_A) \rightarrow \bar{u}_{\updownarrow z}(p'_A)\,\gamma_0\,S^\dagger\,\gamma_0 = \bar{u}_{\updownarrow z}(p'_{B}).
\end{eqnarray}

From here we can construct typical helicity matrix elements 
in the IMF and determine what they should look like in the Breit frame. 
For the helicity non-flip in the IMF, which involves transitions
between two states with spin aligned along the $z$ axis, we have
\begin{equation}
\langle f'_{B}| J^+|i'_{B}\rangle = 2 p_{B}^+ F_1,
\end{equation}
which corresponds to the matrix elements in the Breit frame,
\begin{eqnarray}
\langle f'_{A} | J^0| i'_{A} \rangle &=&2 M \frac{M}{E_0} G_E\nonumber \\
\langle f'_{A} | J^1| i'_{A} \rangle &=&0\nonumber \\
\langle f'_{A} | J^2| i'_{A} \rangle &=&-iQG_M \\
\langle f'_{A} | J^3| i'_{A} \rangle &=&\frac{Q^2}{2E_0} G_M.
\end{eqnarray}
If we consider the helicity flip element in the IMF, we have
\begin{equation}
\langle f''_{B}|J^+|i''_{B}\rangle = p_{B}^+\frac{Q}{M}F_2,
\end{equation}
which corresponds to the matrix elements in the Breit frame,   
\begin{eqnarray}
\langle f''_{A} | J^0| i''_{A}\rangle &=&- M \frac{Q}{E_0} G_E \nonumber \\ 
\langle f''_{A} | J^1| i''_{A}\rangle &=&0 \nonumber \\
\langle f''_{A} | J^2| i''_{A}\rangle &=&0 \\ 
\langle f''_{A} | J^3| i''_{A}\rangle &=& M \frac{Q}{E_0} G_M. \nonumber 
\end{eqnarray}
\section {Discussion}

The key result that we obtain is that a Breit frame matrix element of $J^0$ yielding $G_E$, \eq{second}, is converted via a boost, \eq{boost}, to a matrix element of $J^+$ that yields $F_1$, \eq{jplus}, in the infinite momentum frame. The use of the infinite momentum frame, along with the Drell-Yan condition, $q^+=0$, allows the extraction of a transverse density as the two-dimensional Fourier transform of $F_1$ \cite{soper1,mbimpact,diehl2,MillerNeutron,Carlson:2007xd}.
The present result establishes a connection between the standard Breit frame procedure involving $G_E$ and the infinite momentum frame procedure involving $F_1$, which is related to the transverse density. Future work will be concerned with determining whether or not a connection between the rest frame charge density and the transverse density can be established.  

\section*{ACKNOWLEDGMENTS}
We thank the U.S. DOE (FG02-97ER41014)
for partial support of this work.


\begin{thebibliography} {99}

\bibitem{Gao:2003ag}
  H.~y.~Gao,
  Int.\ J.\ Mod.\ Phys.\  E {\bf 12}, 1 (2003); {\bf 12}, 567(E) (2003).
\bibitem{HydeWright:2004gh}
  C.~E.~Hyde and K.~de Jager,
  Ann.\ Rev.\ Nucl.\ Part.\ Sci.\  {\bf 54}, 217 (2004). 
\bibitem{Perdrisat:2006hj}
  C.~F.~Perdrisat, V.~Punjabi and M.~Vanderhaeghen,
  Prog.\ Part.\ Nucl.\ Phys.\  {\bf 59}, 694 (2007). 
\bibitem{Arrington:2006zm}
  J.~Arrington, C.~D.~Roberts and J.~M.~Zanotti,
  J.\ Phys.\ G {\bf 34}, S23 (2007). 

\bibitem{soper1} D.~E.~Soper,
  Phys.\ Rev.\  D {\bf 15}, 1141 (1977).

\bibitem{mbimpact}
  M.~Burkardt,
  Int.\ J.\ Mod.\ Phys.\  A {\bf 18}, 173 (2003).
\bibitem{diehl2} 
  M.~Diehl,
  Eur.\ Phys.\ J.\  C {\bf 25}, 223 (2002); {\bf 31}, 277(E) (2003).

\bibitem{MillerNeutron} G.~A.~Miller,
  Phys.\ Rev.\ Lett.\  {\bf 99}, 112001 (2007).
\bibitem{Carlson:2007xd}
 C.~E.~Carlson and M.~Vanderhaeghen,
  Phys.\ Rev.\ Lett.\  {\bf 100}, 032004 (2008).

\bibitem{Bjorken}
J. D. Bjorken and S. D. Drell, {\it Relativistic Quantum Mechanics}
(McGraw-Hill, New York, 1964).

\bibitem{Kogut:1969xa} J.~B.~Kogut and D.~E.~Soper,
  Phys.\ Rev.\  D {\bf 1}, 2901 (1970).
\bibitem{Lepage:1980fj}G.~P.~Lepage and S.~J.~Brodsky,
  Phys.\ Rev.\  D {\bf 22}, 2157 (1980).
\bibitem{Frederico:1991vb} T.~Frederico, E.~M.~Henley and G.~A.~Miller,
  Nucl.\ Phys.\  A {\bf 533}, 617 (1991).
\bibitem{ls} L.~Susskind, Phys. \ Rev.\ {\bf 165}, 1535 (1968); Phys. \ Rev.\ {\bf 165}, 1547 (1968).

\end{thebibliography}
\end{document}